\documentclass[conference, compsocconf]{IEEEtran}
\IEEEoverridecommandlockouts
\usepackage{cite}
\usepackage{listings}
\usepackage{amssymb}
\usepackage{bm}
\usepackage{amsmath,amssymb,amsfonts}
\usepackage{graphicx}
\usepackage{textcomp}
\usepackage{xcolor}
\usepackage{booktabs}
\usepackage{url}

\begin{document}

\title{Layer-wise MoE Routing Locality under Shared-Prefix Code Generation: Token-Identity Decomposition and Compile-Equivalent Fork Redundancy}

\author{
\IEEEauthorblockN{Shun-ichiro Hayashi\textsuperscript{1},
Daichi Mukunoki\textsuperscript{2},
Tetsuya Hoshino\textsuperscript{2},
Takahiro Katagiri\textsuperscript{2}}
\IEEEauthorblockA{\textsuperscript{1}Graduate School of Informatics, Nagoya University, Japan\\
\textsuperscript{2}Information Technology Center, Nagoya University, Japan\\
hayashi@hpc.itc.nagoya-u.ac.jp,
\{mukunoki, hoshino, katagiri\}@cc.nagoya-u.ac.jp}
}

\maketitle

\begin{abstract}
In LLM-based code generation, multiple code candidates are often generated
in parallel from the same prompt---for example, in best-of-$N$ sampling or
multi-candidate code completion.
These requests can share KV caches through a common prefix, yet the extent
to which their Mixture-of-Experts (MoE) expert routing overlaps, and how
this overlap varies across layers, remains insufficiently understood.
We study Qwen3.5-35B-A3B-FP8 (256 routed experts, top-8) by performing
tree-search-based branching generation from a shared prefix (851 completed
codes, temperature 0.7) and analyzing the results with a compiler-output-based
alignment (\texttt{gcc -S -O0} assembly) that controls for token-identity
confounds.
Our findings are threefold:
(1)~At positions where both sequences generated the same token, Jaccard
similarity reaches 0.649 (40$\times$ random), while even at positions
with different tokens it remains 0.175 (11$\times$ random).
(2)~A layer-wise decomposition reveals a crossing pattern: same-token
routing similarity exceeds different-token similarity across all layers,
but dips in the middle layers (L14--20), while different-token similarity
peaks in the middle layers at 14$\times$ random.
(3)~In tree-search code generation, 67\% of successfully compiled codes
concentrate in the top three assembly-equivalent groups, and 99.6\% of
within-group differences consist of comments and blank lines.
We show that diversity in top-$P$ search, including beam search,
poses a significant challenge.
These results refine the ``context-independent routing'' claim of prior
work through layer-wise decomposition and suggest opportunities for
improving search efficiency in LLM code generation.
\end{abstract}

\section{Introduction}
\label{sec:intro}

In LLM-based code generation, multiple code candidates are often generated
in parallel from the same prompt.
Multi-candidate code completion and best-of-$N$ sampling for quality
improvement are typical examples.
Because these requests share a common prefix (prompt plus shared context),
KV cache reuse~\cite{sglang} yields high throughput efficiency, and GPU
utilization improves as the number of concurrent requests increases.

However, whether such candidate sets actually exhibit meaningful diversity
has not been sufficiently investigated.
If diversity is ensured, best-of-$N$ search becomes more effective,
opening the possibility of jointly improving code quality and parallel
throughput.
Conversely, if most candidates are essentially identical code, redundant
computation is wasted and the gains of parallelism diminish.

Mixture-of-Experts (MoE) models~\cite{deepseek_v3, qwen35} activate only
a small subset of experts per token.
OpenMoE~\cite{openmoe} aggregated token-ID-level routing distributions
across an entire dataset and reported that expert routing is primarily
determined by token identity rather than context.
However, this analysis
(1)~did not verify whether the same token ID is routed to the same experts
when it appears in different contexts,
(2)~did not decompose routing by layer (reporting only a single layer), and
(3)~did not measure routing similarity when different tokens are generated
from the same prefix.
In this work, we control for the same-prefix/different-token condition
through tree-search-based code generation from a shared prefix and analyze
MoE routing bias along multiple axes---layer, token identity, and
compile equivalence---while quantitatively evaluating code diversity based
on compiler output (assembly).

Regarding MoE inference optimization, neuron-level offloading~\cite{powerinfer},
expert-level offloading~\cite{moe_infinity}, activation
prediction~\cite{fmoe}, and intra-batch expert
sharing~\cite{xshare} have been proposed, but these focus on expert
activation patterns within a single request.
Recently, layer-dependent convergence-divergence patterns in routing have
been reported in natural-language multilingual
settings~\cite{zheng2026unveiling}, but to our knowledge no study has
jointly analyzed routing bias and generation diversity in code generation
tasks.

Our contributions are as follows:
\begin{enumerate}
  \item Measurement of expert-set overlap across branching generation
    requests.
    Prior MoE analyses have focused on intra-request token-level
    activation; we measure the overlap of per-layer expert sets across
    multiple generation sequences branched from a shared prefix via
    tree search, showing that structural overlap exceeding 11$\times$
    random persists in the inter-request direction.
  \item Observation of a layer-dependent crossing structure.
    Decomposing routing by layer reveals a two-phase structure in which
    token identity dominates in the input layers while context dominates
    in the middle layers.
    This result provides material for a layer-wise reinterpretation of
    the ``context-independent'' claim made by OpenMoE based on
    single-layer visualization.
  \item Quantification of code redundancy via compiler output.
    By classifying generated codes according to \texttt{gcc -S -O0}
    assembly output, we quantify that the majority of diversity produced
    by top-$P$ search consists of superficial source-level variation
    such as comments and blank lines, and show that the bottleneck for
    semantic diversity lies in natural-language comments.
  \item Observation of comment-induced algorithm branching.
    Even when comment generation is explicitly prohibited in the prompt,
    differences in leaked reasoning comments trigger divergent dead-code
    structures downstream, demonstrating that diversity in LLM code
    generation depends on natural-language variation.
\end{enumerate}

\section{Related Work}
\label{sec:related}

\subsection{Routing Analysis in MoE Inference}

Prior work on MoE routing bias has primarily targeted short sequences
(tens to hundreds of tokens) within a single request.
MoE-Infinity~\cite{moe_infinity} showed that active experts account for
only 3--20\% of the total under a 512-prompt + 32-generation token setting.
fMoE~\cite{fmoe} analyzed expert activation patterns in offloading
workloads (mean sequence length $\sim$165~tokens).
All of these are intra-request analyses; no study has measured
inter-request routing similarity under parallel generation from a
shared context.

\subsection{Context Dependence of Routing}

OpenMoE~\cite{openmoe} visualized routing patterns over 2048-token
sequences and concluded that routing is ``context-independent.''
However, the quantitative analysis was limited and no rigorous layer-wise
decomposition was performed.
We verify this claim through layer-wise decomposition and show that
while token identity dominates in the input layers, it is contradicted
in the middle layers.

Recently, layer-dependent routing patterns have been reported in
multilingual MoE models.
Zheng et al.~\cite{zheng2026unveiling} observed that routing diverges in
shallow layers (language-specific experts), converges in middle layers
(language-agnostic shared experts), and diverges again in deep layers.
Independently, Nuriyev \& Kulp~\cite{nuriyev2026expert} demonstrated that
expert selections alone carry enough information to reconstruct input
tokens with 91.2\% top-1 accuracy, confirming that routing patterns
encode rich semantic content.
These studies approach routing from different angles (multilingual
structure and information leakage, respectively); our work differs in
that it measures routing similarity across multiple requests branched
from the same context in a monolingual (code generation) setting.
Table~\ref{tab:related_comparison} compares experimental settings with
prior work.

\begin{table*}[t]
\centering
\caption{Comparison of experimental settings with prior work.}
\label{tab:related_comparison}
\begin{tabular}{llrrrrll}
\toprule
Study & Model & \#Experts & Active & Input & Gen & Analysis target & Key feature \\
\midrule
MoE-Infinity~\cite{moe_infinity} & Mixtral et al. & 8--128 & 1--6 & 512 & 32 & Single-seq activation & 3--20\% active \\
fMoE~\cite{fmoe} & Mixtral et al. & 8--60 & 2--4 & \multicolumn{2}{c}{$\sim$165} & Single-seq prediction & Offload opt. \\
OpenMoE~\cite{openmoe} & OpenMoE & 32 & 2 & \multicolumn{2}{c}{2048} & Single-seq visualization & Context-indep. claim \\
Zheng+~\cite{zheng2026unveiling} & Qwen3 et al. & 16--128 & 2--8 & \multicolumn{2}{c}{$\sim$thousands} & Multilingual routing & Conv-div pattern \\
\midrule
\textbf{Ours} & \textbf{Qwen3.5} & \textbf{256} & \textbf{8} & \textbf{98} & \textbf{$\sim$300} & \textbf{Cross-req pairwise} & layer$\times$tok$\times$O0 decomp. \\
\bottomrule
\end{tabular}
\end{table*}

\subsection{Forward Pass Redundancy in Inference Engines}

SGLang~\cite{sglang} and vLLM share KV caches for common prefixes,
but decode-phase forward passes are executed independently for each
request.
Hydragen~\cite{hydragen} achieved shared attention computation, but
sharing at the MoE layer (FFN/expert computation) has been addressed
only through approximate methods.
XShare~\cite{xshare} constrains intra-batch expert activation as an
approximation, but may incur quality degradation.
Our findings suggest the possibility of non-approximate expert sharing
that exploits structural routing bias.

\section{Experimental Setup}
\label{sec:setup}

This study investigates MoE routing characteristics when generating
multiple code candidates in parallel from the same prompt, through the
following experiments:
\begin{enumerate}
  \item Measure expert routing overlap between codes branched from
    a shared prefix via tree-search-based generation.
  \item Control for token-identity confounds (same-tok/diff-tok
    separation) to elucidate the layer-wise routing locality structure.
  \item Quantify generation diversity and redundancy through alignment
    based on compiler output with optimization suppressed
    (\texttt{gcc -S -O0}).
\end{enumerate}

\subsection{Computing Environment}
Experiments were conducted on a single NVIDIA GH200 Grace Hopper
Superchip node (NVIDIA H100 GPU, 96\,GB HBM3 + Grace CPU, 120\,GB
LPDDR5X, NVLink-C2C 900\,GB/s).
The OS was RHEL~9.4 (aarch64) with \texttt{gcc~11.4.1}.

\subsection{Model and Inference Engine}

We used Qwen3.5-35B-A3B-FP8~\cite{qwen35}, a hybrid GDN (Gated Delta
Net) architecture with 40~layers: 30~GDN layers (linear attention)
and 10~full-attention layers (layer indices $4n{+}3$,
$n=0,\ldots,9$).
The MoE configuration is 256~routed experts + 1~shared expert with
top-8 routing.
The shared expert is always active and does not participate in
routing; it is therefore excluded from our routing data.

We used SGLang~0.5.9 as the inference engine.
Generation requests were sent directly to SGLang's native raw text
completion endpoint \texttt{/generate}, rather than the
OpenAI-compatible \texttt{/v1/chat/completions}.
We chose \texttt{/generate} because the latter enables Qwen3.5's
\texttt{<think>} block by default, which is incompatible with the
manual prefix control (\S\ref{sec:tree_search}, thinking skip) and
routing metadata retrieval required by this study.
The server was launched with the
\texttt{--enable-return-routed-experts} flag, and per-token
routed expert IDs were extracted from the response's
\texttt{meta\_info["routed\_experts"]} (a Base64-encoded
\texttt{int32} array).

\subsection{Tree-Search-Based Branching Generation}
\label{sec:tree_search}

C-language code was generated using the following prompt:

\begin{quote}
\small
\texttt{[System]} You are a helpful coding assistant.
Write clean C code.\\
\texttt{[User]} Write a C function that sorts an integer array.
Compiler: gcc (Linux). Standard library only.
The input has 1,000,000 elements, approximately 80\% pre-sorted
with 20\% random insertions. Single-threaded.
Do not include comments. Output code only.
\end{quote}

The key parameters are: target cumulative probability
$P_\mathrm{target}=0.40$, maximum forks per decision point $K=20$,
timeout 30~minutes, and maximum concurrent generations 50.
Sampling uses temperature $T{=}0.7$, top-$p{=}0.95$,
top-$k{=}20$, and maximum generation length 2048~tokens.
Recent reasoning-capable LLMs, including the Qwen3.5 family, generate
a reasoning process inside \texttt{<think>...</think>} tags before the
main response by default.
The official Qwen3 chat template~\cite{qwen35} inserts an empty
\texttt{<think>$\backslash$n$\backslash$n</think>$\backslash$n$\backslash$n}
at the start of the assistant turn when the
\texttt{enable\_thinking=false} argument is given, causing the model
to interpret the thinking block as ``already completed'' and begin
generating the main response directly.
Because our requests are sent via the \texttt{/generate} endpoint,
which does not apply the chat template, we manually inject the same
empty tag sequence at the end of the prefix to reproduce this behavior
(hereafter called \textit{thinking skip}).
This setting prevents thinking tokens from contaminating the routing
measurements and suppresses generation-length inflation; side effects
are discussed in \S\ref{sec:explosion}.
The tree is expanded as follows:
\begin{enumerate}
  \item Generate the root node (until the model emits the
    end-of-response token \texttt{</function>}).
  \item At every decode step of a completed node, detect decision
    points (DPs) where the top-30 logprobs yield cumulative probability
    $\leq P_\mathrm{target}$ and at least two candidates.
  \item Fork up to $K$ children for each candidate token at each DP.
  \item Generate forked nodes asynchronously; scan DPs immediately
    upon completion.
  \item Repeat until all DPs are exhausted or timeout is reached.
\end{enumerate}

The tree at $P{=}P_\mathrm{target}{=}0.40$ diverged
(branching ratio $\approx 6$), and exhaustive search did not complete
within the 30-minute timeout.
The 851~completed codes that reached \texttt{</function>}
(\texttt{finish\_reason=stop}) at timeout were used for analysis;
4,870~truncated codes and 24~errors were excluded.
For each completed code, we recorded token IDs, top-30 logprobs, and
routed expert IDs for all 40~layers $\times$ top-8.

\noindent\textbf{Sibling pairs.}
In the tree structure, if decode step~$d$ of a completed code~$A$
is a decision point, a completed code~$B$ generated by selecting a
different candidate token at that step is called a \textbf{sibling}
of~$A$.
A sibling pair $(A, B)$ shares the entire context from the prompt
through step~$d$ of~$A$, differing only from step $d{+}1$ onward.
This sharing ensures that step-aligned comparison between siblings is
exact from step~$d$ onward.
Twenty sibling pairs were obtained in this experiment.

\subsection{Compile-Equivalence-Based Alignment}
\label{sec:alignment}

The nodes obtained by tree search differ in token sequence length due
to variation in comments, whitespace, and newlines.
Direct comparison by step number (step-aligned) causes misalignment
to accumulate from comment insertions, resulting in comparison of
unrelated code positions.

We generated assembly with \texttt{gcc -S -O0} (aarch64, on GH200)
and classified nodes producing identical assembly into
\textbf{O0-equivalent groups}, which serve as the analysis units
hereafter.
Of 851~completed codes, 691 compiled successfully (81\%), forming
189~O0-equivalent groups (top three groups: 219, 201, and 40~nodes).

Within an O0-equivalent group the code bodies are identical strings;
we used Python's \texttt{difflib.SequenceMatcher} for token-sequence
alignment to identify corresponding token positions, classifying them
as same-token (same-tok) or different-token (diff-tok) positions.

\subsection{Metrics}

For each layer, we compute the Jaccard similarity
$J = |A \cap B| / |A \cup B|$ of the expert sets (8~experts each)
between two nodes, reporting either the all-layer average or
layer-specific values.
The theoretical Jaccard value when randomly selecting 8 out of
256~routed experts is approximately 0.016.

The effective number of experts is defined as $2^H$, where $H$ is the
Shannon entropy of the expert selection distribution.

\section{Results}
\label{sec:results}

\subsection{Overview of Routing Locality}
\label{sec:overview}

In the following sections, Jaccard similarity is measured using
token-position pairs under four conditions:
\begin{itemize}
  \item W-same: positions within an O0-equivalent group where both
    nodes generated the same token.
  \item W-diff: positions within an O0-equivalent group where both
    nodes generated different tokens.
  \item B-same: positions between different O0-equivalent groups where
    both nodes generated the same token.
  \item B-diff: positions between different O0-equivalent groups where
    both nodes generated different tokens.
\end{itemize}

Table~\ref{tab:full_table} shows Jaccard similarity by layer under
these four conditions.

\begin{table}[t]
\centering
\caption{Layer-wise Jaccard similarity ($P$=0.40, $n$=851 completed
  codes). W=Within O0-equivalent (aligned), B=Between O0 groups
  (aligned). Random baseline 0.016.}
\label{tab:full_table}
\begin{tabular}{lrrrr}
\toprule
Layer & W-same & W-diff & B-same & B-diff \\
\midrule
0     & 0.828  & 0.087  & 0.709  & 0.101  \\
4     & 0.674  & 0.122  & 0.513  & 0.103  \\
8     & 0.664  & 0.177  & 0.495  & 0.190  \\
14    & 0.618  & 0.223  & 0.485  & 0.223  \\
20    & 0.597  & 0.212  & 0.437  & 0.194  \\
26    & 0.575  & 0.194  & 0.433  & 0.188  \\
34    & 0.666  & 0.116  & 0.483  & 0.156  \\
38    & 0.616  & 0.207  & 0.401  & 0.118  \\
\midrule
Mean  & 0.649  & 0.179  & 0.483  & 0.175  \\
\bottomrule
\end{tabular}
\end{table}

\begin{figure}[t]
    \centering
    \includegraphics[width=\columnwidth]{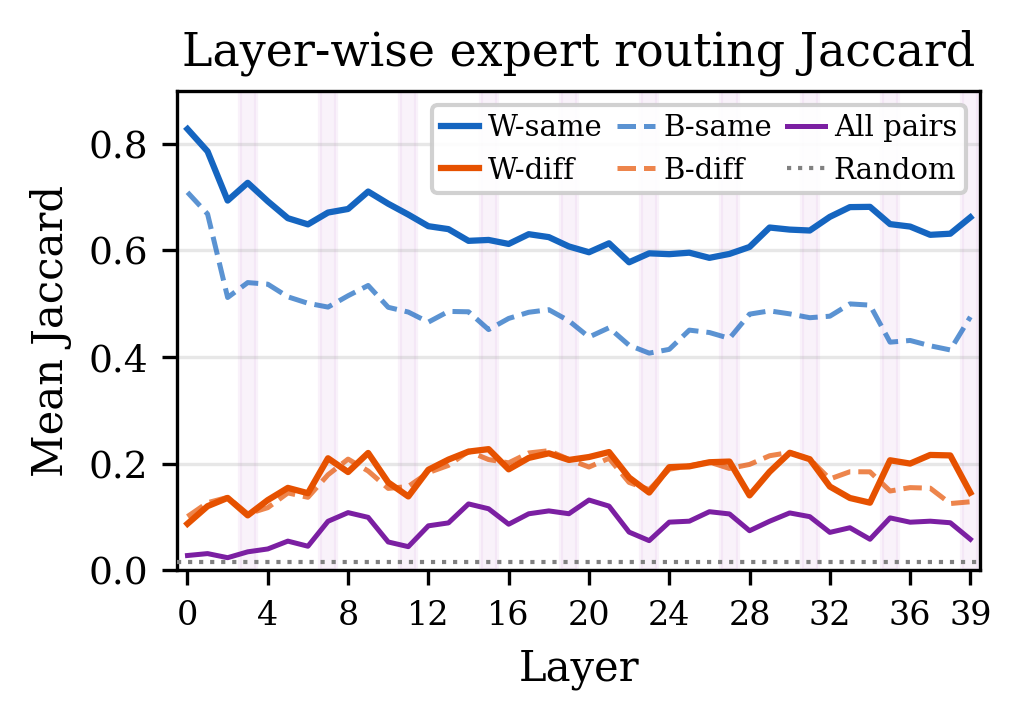}
    \caption{Layer-wise Jaccard similarity crossing pattern
      ($n$=851 completed codes).
      W=Within O0-equivalent (solid), B=Between O0 groups (dashed),
      All pairs=all-pair aggregate (purple).
      Same-token (blue) decreases in the middle layers;
      different-token (orange) peaks in the middle layers.
      Pink bands indicate full-attention layers ($4n{+}3$).
      Random baseline 0.016 (gray dotted).}
    \label{fig:crossing}
\end{figure}

Jaccard values under all conditions substantially exceed the random
baseline (0.016).
Even under the strictest condition (B-diff: different tokens between
different O0 groups), the value is 0.175 (11$\times$ random),
confirming that the shared prefix induces structural routing bias.
Figure~\ref{fig:crossing} plots these four conditions by layer.
The all-pair aggregate (purple line in Fig.~\ref{fig:crossing})
nearly overlaps the diff-tok line because diff-tok pairs dominate.

\subsection{Layer-Dependent Crossing Pattern}
\label{sec:crossing}

From Fig.~\ref{fig:crossing}, same-token and different-token pairs
exhibit contrasting patterns:

\begin{itemize}
  \item Same-tok (blue): maximum at the input layer
    (L0 $\approx$ 0.8), dipping in the middle layers
    (L14--20 $\approx$ 0.6), with partial recovery in the deep layers.
  \item Diff-tok (orange): minimum at the input layer
    (L0 $\approx$ 0.09, low compared to the middle layers), peaking
    in the middle layers (L14 $\approx$ 0.22, 14$\times$ random),
    then declining in the deep layers.
\end{itemize}

Within (solid) and Between (dashed) show the same trend, but for
same-tok, Within $>$ Between (code structure is reflected in routing),
whereas for diff-tok, Within $\approx$ Between (code structure has
little influence).

\begin{figure}[t]
    \centering
    \includegraphics[width=\columnwidth]{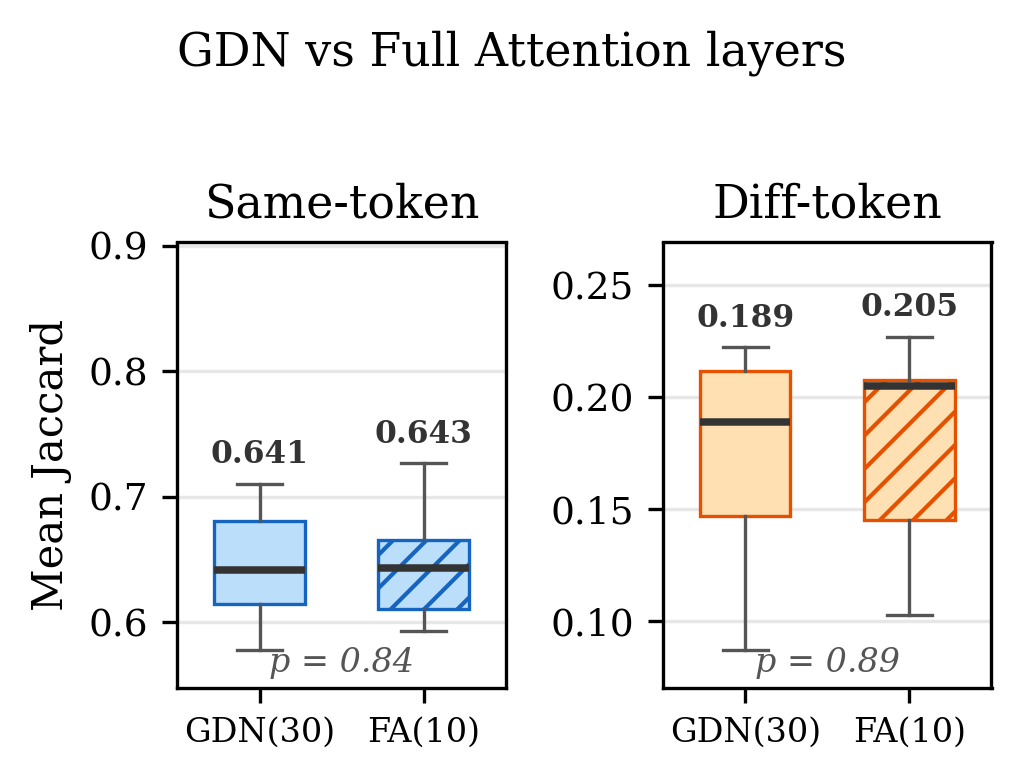}
    \caption{Jaccard similarity distribution for MoE layers preceded
      by GDN (30 layers) vs.\ full attention (10 layers).
      No significant difference ($p > 0.8$).
      FA boxes are hatched.}
    \label{fig:gdn_fa}
\end{figure}

No significant difference in routing characteristics was observed
between GDN layers (30/40 layers) and full-attention layers
(10/40 layers)
(same-tok: $p$=0.84, diff-tok: $p$=0.89, Mann-Whitney U test;
Fig.~\ref{fig:gdn_fa}).
What dominates routing characteristics is layer position, not the type
of attention mechanism immediately preceding the MoE layer.

\subsection{L14 Expert Co-Activation Structure}
\label{sec:coact_l14}

\begin{figure}[t]
    \centering
    \includegraphics[width=\columnwidth]{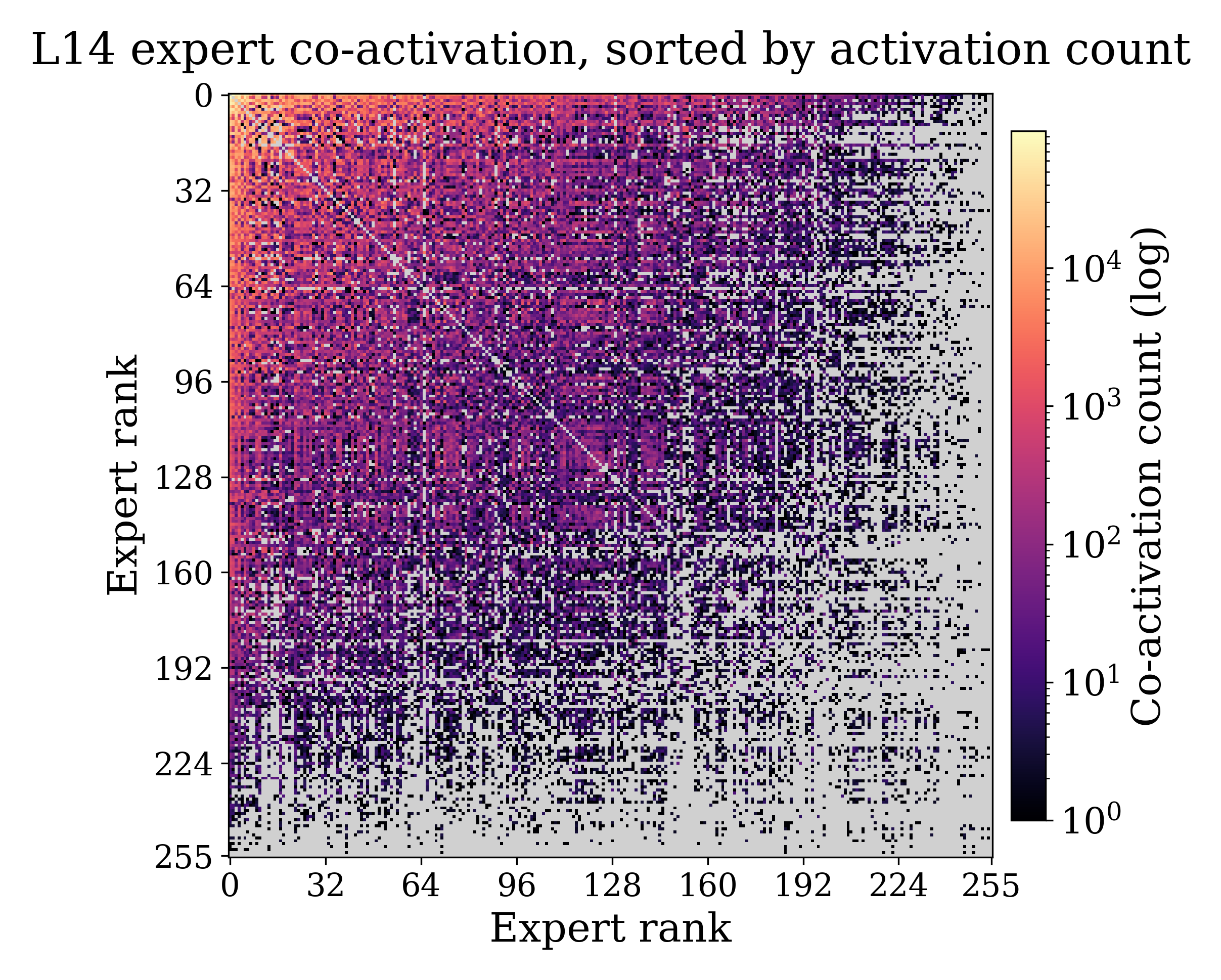}
    \caption{$256 \times 256$ expert co-activation matrix at L14
      ($n$=851 completed codes).
      Each cell $(i,j)$ shows the number of token positions where
      experts $i$ and $j$ were both selected in top-8 (log scale).
      Axes are sorted by L14 activation count in descending order;
      diagonal and zero-count cells are masked in gray.}
    \label{fig:coact_l14}
\end{figure}

Jaccard similarity is a relative measure between expert sets and does
not convey how many times individual experts are co-selected in top-8
with other experts---i.e., absolute co-activation counts.
In this section, we select a specific layer and directly observe
co-activation counts.
As a representative, we choose L14, the layer identified in
\S\ref{sec:crossing} as the diff-tok Jaccard peak.
Figure~\ref{fig:coact_l14} shows the $256 \times 256$ expert
co-activation matrix at L14, sorted in descending order of activation
count (rank~0 corresponds to the most-activated expert).

At the top-left corner (rank~0--4), the five most frequently
activated experts (id=132, 248, 44, 173, 18) are placed; at the
bottom-right (rank~255), expert id=11 is positioned.
Of the total 244,216 token positions, id=132 appears in 163,756
(67\%) of L14 top-8 selections.
Expert id=11 was never activated at L14 in this experiment.

Of all $\binom{256}{2}=32{,}640$ expert pairs, 12,677 pairs (38.8\%)
had zero co-activation, a degree of structured zero-dominance that
cannot arise from random top-8 selection.

The above observations are specific to this experiment (shared-prefix
code generation, Qwen3.5-35B-A3B-FP8, $n$=851) at L14;
generalization to other tasks or MoE models requires further study.
The crossing pattern reported in \S\ref{sec:crossing} (diff-tok
Jaccard peak in the middle layers) can be interpreted as arising
against the backdrop of such hierarchical routing structure in the
middle layers.

\subsection{Routing Overlap Decay after Fork}
\label{sec:decay}

\begin{figure}[t]
    \centering
    \includegraphics[width=\columnwidth]{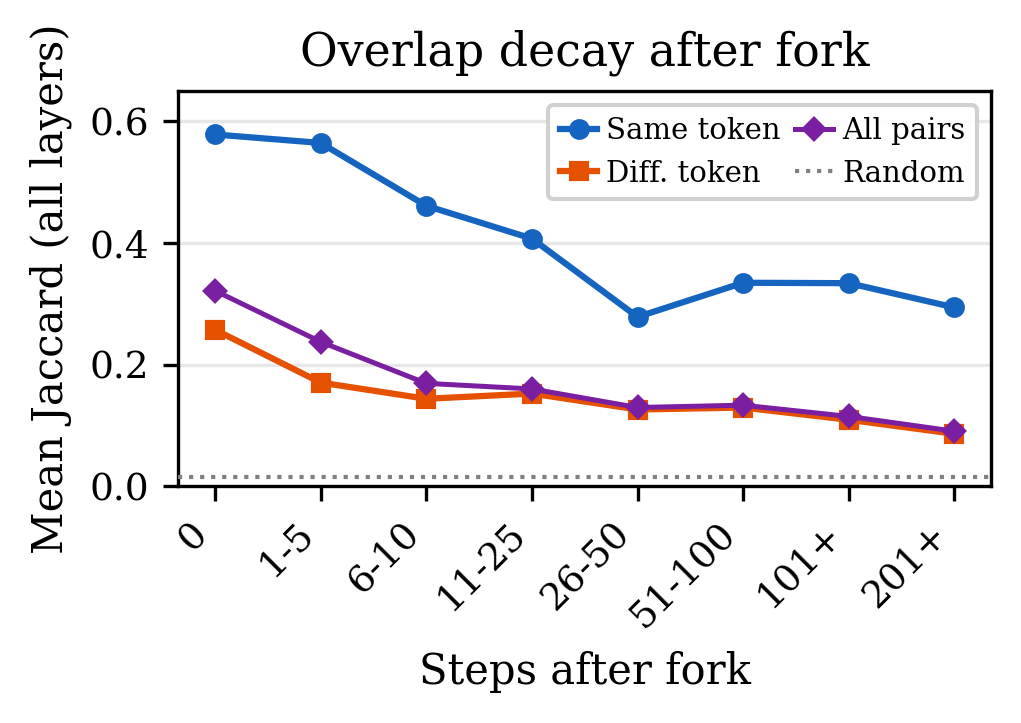}
    \caption{Routing overlap decay after fork (same-branch\_step
      sibling pairs, $n$=851 completed codes).
      Both same-token pairs (blue) and different-token pairs (orange)
      remain above the random baseline (red dashed) throughout.
      Gray bars show the same-token ratio (right axis).}
    \label{fig:decay}
\end{figure}

Figure~\ref{fig:decay} shows the decay of routing overlap with decode
steps for sibling pairs (20 pairs) generated from the same branching
point.

Same-token Jaccard decreases from 0.578 immediately after the fork to
0.294 after 200~steps, but remains at 18$\times$ the random baseline.
Different-token Jaccard decreases from 0.257 to 0.086, stabilizing at
approximately 5$\times$ random.
The same-token ratio drops sharply from 20\% immediately after the
fork to 2\% at distant steps, indicating rapid context differentiation
after branching.

\subsection{Token-Type Dependence}
\label{sec:token_type}

Table~\ref{tab:token_type} shows Jaccard similarity by token type
within O0-equivalent groups.

\begin{table}[t]
\centering
\caption{Jaccard similarity by token type (Within O0-equivalent,
  all-layer average).
  $n$ is the cumulative count of token positions satisfying each
  condition across pair traversals (same-tok: positions where both
  nodes selected the same token; diff-tok: positions where they
  selected different tokens).}
\label{tab:token_type}
\begin{tabular}{lrrrr}
\toprule
Type & same-tok $J$ & $n$ & diff-tok $J$ & $n$ \\
\midrule
keyword     & 0.685 &    23 & 0.171 &  1579 \\
punctuation & 0.668 &  4856 & 0.152 &   490 \\
identifier  & 0.653 & 11058 & 0.193 &  7112 \\
operator    & 0.627 &   208 & 0.132 &  8835 \\
whitespace  & 0.594 &  7697 & 0.125 &  7409 \\
comment     & 0.553 &    49 & 0.107 &  1336 \\
number      & 0.522 &   428 & 0.095 &   799 \\
\bottomrule
\end{tabular}
\end{table}

For same-tok, keywords (0.685) score highest, indicating that
structurally deterministic tokens exhibit the most consistent routing.
For diff-tok, identifier positions (0.193, 12$\times$ random) score
highest, indicating that context-dependent routing is strongest at
variable-name and function-name positions.
At L14, identifier positions reach 0.235 (15$\times$ random).

Numeric tokens score lowest on both metrics; routing of numeric values
shows the weakest context dependence.

\section{Discussion}
\label{sec:discussion}

\subsection{Interpretation of the Crossing Pattern}

The crossing pattern observed in this study (same-tok dipping in the
middle layers while diff-tok rises) likely reflects the process by
which each MoE layer progressively transforms token representations.

In the input layer (L0), token surface form dominates: same tokens are
routed to the same experts (same-tok = 0.828), while different-token
routing similarity is low compared to the middle layers
(diff-tok $\approx$ 0.09, 5--6$\times$ random).
In the middle layers (L14--20), contextual information is
incorporated: even identical tokens may be routed to different experts
when surrounding context differs (same-tok decreases), and conversely,
different tokens sharing sufficient context acquire similar
representations and are routed to similar experts (diff-tok increases).

This interpretation is consistent with the
divergence-convergence-divergence pattern reported by
Zheng et al.~\cite{zheng2026unveiling} in multilingual settings;
a similar layer-wise structure is observed in our monolingual
code-generation setting.
However, this study is limited to code generation, where superficial
variation in newlines, whitespace, and comments dominates branching
(\S\ref{sec:explosion}); natural-language generation may exhibit
different characteristics.
Replication in other domains is left for future work.

\subsection{Refinement of the OpenMoE Claim}

OpenMoE~\cite{openmoe} concluded that ``routing is context-independent
and dominated by token identity.''
Our results refine this claim through layer-wise decomposition:
token identity is dominant in the input layers (L0 diff-tok $\approx$
0.09, low relative to the middle layers), but the middle layers are
clearly context-dependent (L14 diff-tok = 0.223, 14$\times$ random).
This discrepancy may stem from OpenMoE's lack of layer-wise
decomposition.

\subsection{Implications for Expert Offloading}

The existence of routing locality has implications for CPU/GPU expert
offloading in MoE models.
Because context-dependent routing peaks in the middle layers, keeping
a small set of middle-layer experts GPU-resident is particularly
effective for parallel generation from a shared prefix.

Moreover, the fact that diff-tok Jaccard is 0.175 (11$\times$ random)
even between nodes in different O0-equivalent groups indicates that
branched request groups tend to activate similar expert sets, implying
high predictability of expert activation.

\subsection{Comment-Induced Algorithm Branching}

Detailed examination of inter-O0-group differences reveals that the
forced tokens at group-separating branch points are natural-language
text within comments, such as \texttt{standard}, \texttt{merge}, and
\texttt{custom}.
That is, the LLM generating different comment explanations propagated
into differences in subsequent code structure (algorithm selection).

\begin{figure}[t]
    \centering
    \includegraphics[width=\columnwidth]{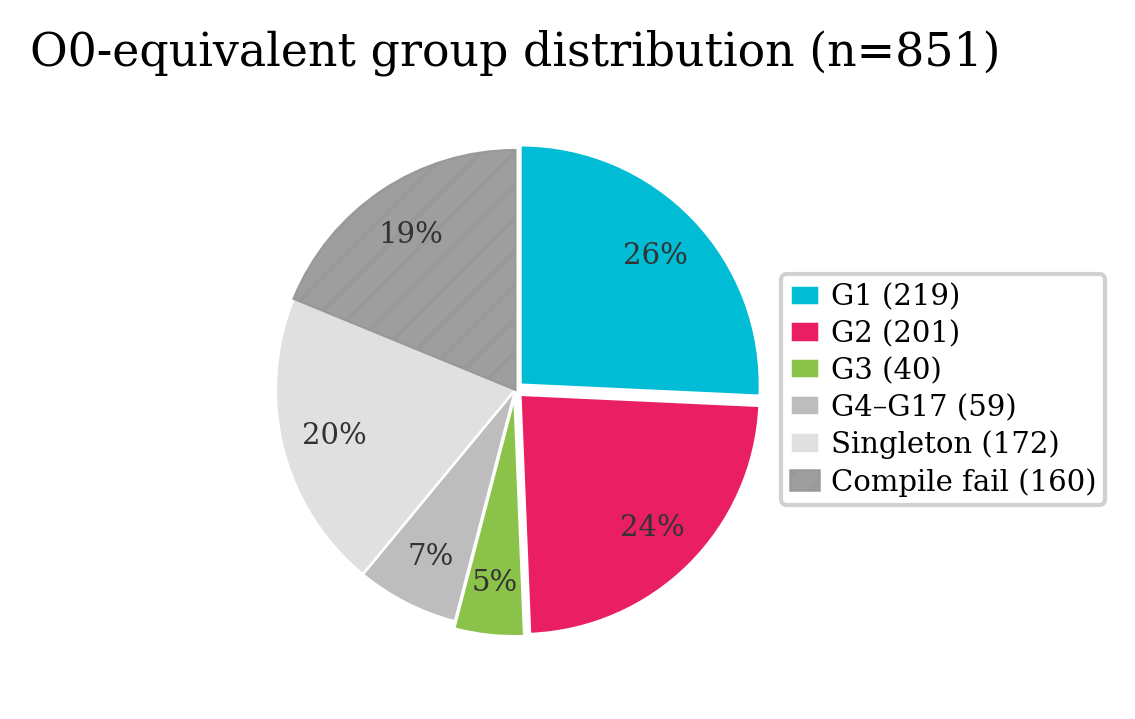}
    \caption{Distribution of O0-equivalent groups ($n$=851 completed
      codes). G1+G2 account for 49\%; the top three groups cover 54\%.}
    \label{fig:o0_dist}
\end{figure}

Figure~\ref{fig:o0_dist} shows the distribution of nodes across
O0-equivalent groups.
Hereafter, the top groups are referred to as G1, G2, G3 (G for group).
G1 (219~nodes, 26\%) and G2 (201~nodes, 24\%) account for 49\% of
the total; adding G3 (40~nodes) brings the coverage to 54\%.
Of the remaining 231~codes, 172~singletons (codes whose O0 assembly
matched no other code) account for 20\%.

Representative code for the three groups is shown in
Listings~\ref{lst:common}--\ref{lst:g3}.
Although the prompt specifies ``80\% pre-sorted, 1M elements,''
the LLM attempts run detection (lines~16--36) but the merge logic
(lines~37--44) becomes unused dead code, ultimately falling back to
\texttt{qsort()} (lines~45--46).
All three groups follow this pattern; differences lie only in the
dead-code portion from line~38 onward.

G1 and G2 are identical through lines~1--44; only the call order of
\texttt{free(buf)} and \texttt{qsort()} in lines~45--46 differs.
G1 (\texttt{free}$\to$\texttt{qsort}) risks use-after-free if
intermediate results remain in \texttt{buf};
G2 (\texttt{qsort}$\to$\texttt{free}) is safer.
G3 differs in variable declarations from line~39 onward
(\texttt{int k} vs.\ \texttt{int len}), but all are dead code and
execution results are identical across the three groups.

\begin{lstlisting}[basicstyle=\ttfamily\scriptsize, frame=single,
  caption={Common portion of the three groups (lines 1--38)},
  label=lst:common]
int compare_ints(const void *a, const void *b) {
    int ia = *(const int *)a;
    int ib = *(const int *)b;
    if (ia < ib) return -1;
    if (ia > ib) return 1;
    return 0;
}
void sort_mixed_array(int *arr, int n) {
    int *buf = (int *)malloc(n * sizeof(int));
    if (!buf) return;
    int *run_start = arr;
    int run_len = 1;
    int i;
    for (i = 1; i < n; i++) {     /* run detection */
        if (arr[i] < arr[i-1]) {
            /* ... copy to buf + partial qsort ... */
            run_len = 1; continue;
        }
        run_len++;
    }
    /* last run also copied to buf + qsort */
    int *dst = arr; int *src = buf;
\end{lstlisting}

\begin{lstlisting}[basicstyle=\ttfamily\scriptsize, frame=single,
  caption={G1 (219): dead code followed by free$\to$qsort},
  label=lst:g1]
    int k = 0;           /* dead code below */
    int run_count = 0;
    int runs[10000];     /* unused */
    @F@free(buf);@/F@           /* G1: free first */
    @q@qsort(arr, n, sizeof(int), compare_ints);@/q@
}
\end{lstlisting}

\begin{lstlisting}[basicstyle=\ttfamily\scriptsize, frame=single,
  caption={G2 (201): dead code followed by qsort$\to$free},
  label=lst:g2]
    int k = 0;           /* same dead code */
    int run_count = 0;
    int runs[10000];
    @Q@qsort(arr, n, sizeof(int), compare_ints);@/Q@
    @f@free(buf);@/f@           /* G2: free after */
}
\end{lstlisting}

\begin{lstlisting}[basicstyle=\ttfamily\scriptsize, frame=single,
  caption={G3 (40): different dead-code variables},
  label=lst:g3]
    @G@int len = 0;@/G@
    @g@int prev = -2147483648;@/g@
    @g@int i_prev = 0; int k = 0;@/g@
    @q@qsort(arr, n, sizeof(int), compare_ints);@/q@
    @f@free(buf);@/f@
}
\end{lstlisting}

Within each group, function names (\texttt{compare\_ints},
\texttt{sort\_mixed\_array}) and variable names are identical across
all nodes, and 99.6\% of differences consist of comments and blank
lines.

These results indicate that in LLM code generation:
(1)~natural-language text in comments triggers dead-code structure
selection,
(2)~variable-name selection for identical code structures is highly
deterministic, and
(3)~even when conditions are explicitly stated in the prompt, the
model tends to converge to a qsort fallback.

\subsection{Computational Explosion of Branching Trees and Whitespace Dominance}
\label{sec:explosion}

In tree search, increasing $P_\mathrm{target}$ causes the branching
tree to diverge exponentially.
The critical threshold at temperature $T$=0.7 is
$P_c \approx 0.36$; at $P$=0.40, 5,745~forks had been generated by
the time 851~completed codes were obtained
(branching ratio $\approx 6$).

The primary driver of this explosion is branching that produces no
difference in execution results, caused by newlines, whitespace, and
comments.
Of the 851~completed codes at $P$=0.40, 691 compiled successfully,
forming 189~O0-equivalent groups; the top three groups (219, 201,
40~nodes) account for 67\% of the total.
Within these groups, O0 assembly is identical.
Exhaustive classification of differences within the largest group
(219~nodes) shows that 88.4\% are comment-wording variations and
11.2\% are blank-line presence/absence, yielding 99.6\% comments or
blank lines overall.
Notably, comments are generated despite the prompt specifying
``Do not include comments.''
This is attributed to the thinking skip causing the model to output
design reasoning---which would normally appear in the
\texttt{<think>} block---as post-code comments (12--30~lines).
Execution-code differences account for only 0.04\%; virtually
identical code was generated in large quantities.

Listing~\ref{lst:thinking_leak} shows a typical example.
A node with 3~comment lines (57~lines total) and a node with
112~comment lines (168~lines total) become identical 47-line code
after comment removal.
The 112-line version contains thinking-style self-questioning such as
``Actually, the most efficient...'',
``Let's just use qsort.'',
``Is there a better way?'', and
``Okay, the most practical solution...'',
suggesting it is substitute output for the \texttt{<think>} block.

\begin{lstlisting}[basicstyle=\ttfamily\scriptsize, frame=single,
  caption={Typical comment difference within the same O0 group (G1)},
  label=lst:thinking_leak]
=== Node A (57 lines = code 47 + comment 3) ===
  ...
  // Simplified approach: Since 80% is sorted,
  // just use stdlib qsort.
  @F@free(buf);@/F@
  @q@qsort(arr, n, sizeof(int), compare_ints);@/q@

=== Node B (168 lines = code 47 + comment 112) ===
  ...
  // Actually, the most efficient...is to use qsort
  // But if we want to be smarter:
  // 1. Find the unsorted segments.
  // Let's just use qsort.
  // Is there a better way?
  // Okay, the most practical solution... is to use
  // Let's just use qsort.
  @F@free(buf);@/F@
  @q@qsort(arr, n, sizeof(int), compare_ints);@/q@

=== After comment removal: IDENTICAL (47 lines) ===
\end{lstlisting}

This problem also arises in beam search: many of the $K$ beam slots
are consumed by whitespace and newline variants, preventing allocation
of beam budget to semantically distinct code branches.
AST normalization or compile-equivalence-based beam pruning could
serve as effective countermeasures.

\subsection{Limitations and Future Work}

(1)~This experiment uses only one C code generation task; generality
to natural language or other programming languages is unverified.
(2)~The tree search at $P_\mathrm{target}$=0.40 diverged at
temperature $T$=0.7 (branching ratio $\approx 6$) and did not reach
exhaustive exploration.
The 851~completed codes reached \texttt{</function>} within the
30-minute timeout and represent partial exploration.
The critical threshold is estimated at $P_c \approx 0.36$.
(3)~Alignment coverage is approximately 26\%, and token-sequence
overlap between nodes at different depths is limited.
(4)~Results are from a single model; reproducibility on different MoE
architectures is left for future work.
(5)~No verification has been performed translating routing locality
into actual offloading throughput.
(6)~Thinking skip (\texttt{<think>} block emptied before generation)
was applied, a special condition under which the model's reasoning
process leaks into comments.
Under standard generation with thinking enabled, reasoning leakage
into comments may be suppressed, potentially improving branching
diversity.
However, enabling thinking causes each request to generate a distinct
reasoning path almost immediately, shortening the shared prefix and
reducing KV cache sharing---a trade-off.

\section{Conclusion}
\label{sec:conclusion}

We systematically measured MoE expert routing bias across multiple
generation sequences branched from a shared prefix via tree-search-based
generation and compiler-output-based alignment.

Inter-request expert-set overlap reaches 11$\times$ random even at
positions where different tokens were generated, demonstrating that
the shared prefix leaves a strong correlation in downstream expert
activation.

Routing locality exhibits a layer-dependent crossing structure: token
identity dominates in the input layers, while context-dependent routing
peaks in the middle layers (diff-tok: 14$\times$ random).
This result provides material for a layer-wise reinterpretation of the
``context-independent routing'' claim of OpenMoE.

In tree-search code generation, 67\% of successfully compiled codes
concentrate in the top three O0-equivalent groups, and 99.6\% of
within-group differences are execution-equivalent redundancy caused by
comments and blank lines.
We quantitatively demonstrated that in top-$P$ search and beam search,
resources are not allocated to semantically distinct code branches.
Runtime AST-equivalence-based pruning could serve as an effective
countermeasure.

Furthermore, we observed that differences in leaked reasoning comments
trigger divergent dead-code structures downstream, even when comment
generation is explicitly prohibited in the prompt.
This demonstrates that diversity in LLM code generation depends on
natural-language variation.

These findings provide implications for MoE expert offloading
strategies and highlight the challenges of ensuring diversity in LLM
code generation.

\section*{Acknowledgment}
This work was supported by the Joint Usage/Research Center for
Interdisciplinary Large-scale Information Infrastructures (JHPCN) and
the High-Performance Computing Infrastructure (HPCI) under Project IDs
jh250015 and jh260017.
It was also partially supported by JSPS KAKENHI Grant Number JP24K02945.
In addition, this work was supported by the JST Research and Development
Program for Next-generation Edge AI Semiconductors (Grant Number JPMJES2511).

\bibliographystyle{IEEEtran}
\bibliography{moe_routing}

\end{document}